\documentclass[11pt]{article}
\usepackage{epsf}

\title{\bf On the inclusive gluon production in the Lipatov effective
action formalism}
\author{M.A.Braun, M.Yu.Salykin and M.I.Vyazovsky\\
Dep. of High Energy physics,
 Saint-Petersburg State University,\\
198504 S.Petersburg, Russia}
\def\beq{\begin{equation}}
\def\eeq{\end{equation}}

\def\ep{\epsilon^*_\perp}

\def\noi{\noindent}
\def\bl{{\bf l}}
\def\bp{{\bf p}}
\def\bq{{\bf q}}
\def\bbl{{\bf L}}
\def\bbb{{\bf B}}
\def\be{{\bf e}}

\def\lra{\leftrightarrow}
\begin{document}
\maketitle
\medskip

\noi{\Large\bf Abstract}

The process of gluon production in quark-nucleus collisions is studied
in the framework of Lipatov's
effective action formalism. The relevant simplifications and rules for
longitudinal integrations are discussed in detail.
The results obtained with the help of these rules correspond to the
purely transversal formalism based on effective vertices in the
transversal space.


\section{Introduction}

In the framework of the perturbative QCD, in the Regge kinematics,
particle interaction can be described by the exchange of reggeized gluons
which emit and absorb real gluons and also may split into several
reggeized gluons. Emission of real gluons from  reggeized gluons is
described by vertices introduced in  ~\cite{bfkl} for non-split
reggeons ("Lipatov vertices") and in ~\cite{bartels} for split reggeons
("Bartels vertices").
Originally both type of vertices were calculated directly
from the relevant simple Feynman diagrams in the Regge kinematics.
Later a powerful effective action formalism was proposed by L.N.Lipatov
~\cite{lipatov}, which considers reggeized and normal gluons as independent
entities from the start and thus allows to calculate all QCD diagrams
in the Regge kinematics automatically and in a systematic and self-consistent
way. However the resulting expressions are 4-dimensional and need
reduction to the final 2-dimensional transverse form. This reduction is
trivial  for tree diagrams but becomes less trivial for diagrams
with loops.

In the paper of two co-authors of the present paper (M.A.B. and M.I.V.)
~\cite{bravyaz} it was demonstrated that the diffractive
amplitude for the production of a real gluon calculated by means of the
Lipatov effective action and based on the Reggeon $\to$ one or two
Reggeons and Particle (R$\to$R(R)P) vertices, after integration over
longitudinal variables, goes over into the transversal expression obtained
via the Lipatov and Bartels vertices ("BFKL-Bartels formalism").
However in the process of reduction to the transverse form a certain
prescription had to be used to give sense to divergent integrals.

The total inclusive cross-section off the nucleus, apart from the
diffractive
contribution, contains a contribution from intermediate inelastic
and possibly coloured  states. This latter contribution has a
structure different from the studied diffractive one. In particular,
for the double scattering and  in
the lowest order, with which we limit ourselves here, a part of it is
constructed as a square modulus of  tree production amplitudes with the same
R$\to$R(R)P vertices, so that the loops only
appear at the stage of the formation of the cross-section itself.
Correspondingly the longitudinal integrals which appear have
a more complicated structure as compared to the diffractive contribution.
In this paper we study this part of the non-diffractive contribution.
We show that
after integration over longitudinal variables it also coincides with the
result obtained with the help of Lipatov and Bartels vertices directly in
the transverse space ~\cite{braun}
provided a certain part of the R$\to$RRP vertex is dropped. This part was
demonstrated to be absent in  a particular kinematics, relevant for
the inclusive cross-section ~\cite{new}. Our result confirms that the
restoration of the unitarity contribution from the triple discontinuity
of the amplitude, which can be proven for the total cross-section, remains
valid also for the inclusive cross-section.

Note that a second part of the non-diffractive contribution corresponds
to the product of  tree amplitudes and amplitudes with a loop. The study of
this 'single cut' contribution requires knowledge of a more complicated
R$\to$RRRP vertex and is postponed for future publications.

In our derivation we use a simplified picture, in which both incident
and target particles are quarks. Also, we restrict ourselves to
the double scattering and lowest non-trivial order of perturbation expansion.

\section{Inclusive production off the nucleus}

\begin{figure}[h]
\leavevmode \centering{\epsfysize=0.3\textheight\epsfbox{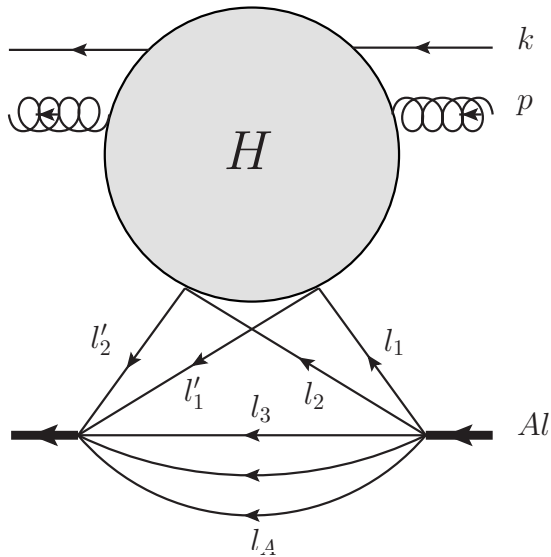}}
\caption{Inclusive production off the nucleus}
\label{fig1}
\end{figure}

We start with reviewing the Glauber picture of gluon production
off the nucleus of atomic number $A$
coming from the double scattering on the nuclear components.
This derivation closely follows the original one in~\cite{gribov}
(see also a later presentation in ~\cite{CK}).
There one can find a detailed description of the separation
of the nuclear part, which is briefly repeated in the following.
Our specific goal is to pass to light-cone variables and c.m. system
for the high-energy part and find the kinematical region of momenta
relevant for the calculation of the inclusive cross-section.
The corresponding diagram is shown in Fig. \ref{fig1}. The blob $H$
corresponds to the high-energy part. The part
belonging to the nucleus is indicated by nucleon propagators attached
to the high-energy blob. For the inclusive cross-section  the blob $H$
is to be cut in the center
but this is not shown, since in fact we have to take its imaginary part
above the cut corresponding to the variable
$M^2=(k+l_1+l_2-p)^2 >0$ having the meaning
of the missing mass squared for gluon production.
So we may start from the blob itself.
Separation of the nuclear part is standardly done in the nuclear rest
system (lab. system), in which the total nucleus 3-momentum $A{\bf l}$
is zero. Individual nuclear momenta have their components
\begin{equation}
l_i=l+\lambda_i,\ \ l'_i=l+\lambda'_i,\ \ l=(m-\epsilon, \vec{0}),
\end{equation}
where $m$ is the
nucleon mass, $\epsilon\to 0$ the binding energy per nucleon and
where $\lambda_i=\lambda'_i$ for $i\geq 3$.
In the nucleus $\lambda_{i0}\sim\epsilon$ and
$\lambda_{iz}\sim\lambda_{i\perp}\sim\sqrt{m\epsilon}$.
So $|\lambda_{i0}|<<|\vec{\lambda}_i|<<l_0$ and similarly for the
primed momenta.

Now we turn to the high-energy blob $H$. It is a Lorenz invariant
function of nuclear momenta depending on $\lambda^2$ and  products
$kl_1, kl_2, k\lambda$ and $pl_1,pl_2,p\lambda$ where
$\lambda=l'_1-l_1=l_2-l'_2$.
In the lab. system we find $kl_1\simeq kl_2\simeq kl$, $pl_1\simeq pl_2
\simeq pl$ and do not depend on nuclear momenta at all. The only dependence
comes from $k\lambda$, $p\lambda$ and $\lambda^2$. Taking into account
that $\lambda_0<<\lambda_z\to 0$
and that $k_{\perp},p_{\perp}<<k_z,p_z\to\infty$ we find that in fact
the only relevant variables are
\begin{equation}
k\lambda=-k_z\lambda_z,\ \ p\lambda=-p_z\lambda_z,
\end{equation}
so that the high-energy part depends only on $\lambda_z$ via $(k\lambda)$
and $(p\lambda)$:
$H=H(\lambda_z)=H(p\lambda,k\lambda)$.

The separation of the nuclear part then easily follows. We have $(A-1)$
independent nuclear momenta, which may be chosen in a different manner.
First we perform integrations over  zero-components, taking as
independent variables $\lambda_{20},\lambda'_{20},\lambda_{i0},\ \ i\geq 3$.
Integration over $\lambda_{i0},\ \ i\geq 3$ is done automatically when we
take the cut, which puts these nucleons on the mass-shell.
The dependence on $\lambda_{20}$ and $\lambda'_{20}$ is assumed to be
contained in four propagators for the active nucleons 1 and 2. The vertex
for the decomposition of the nucleus as a whole into free nucleons is
assumed to be independent of zero-components, which physically corresponds
to absence of essential retardation in the nucleon-nucleon potentials.
Then integration over $\lambda_{20}$ and $\lambda'_{20}$ is done trivially
and gives two denominators, which together with the above mentioned vertex
form the product of nuclear wave functions in the momentum space
\begin{equation}
\frac{1}{2}A(A-1)\frac{(2\pi)^3}{m}
\psi(\bl_1,\bl_2,...\bl_A)\psi^*(\bl'_1,\bl'_2,...\bl'_A),
\label{0a}
\end{equation}
(the coefficient is
easily determined from the comparison with the
baryonic nuclear form-factor at zero transferred momenta).
Integration over the 3-momenta of spectator nucleons 3,...A converts
this into the density matrix for the two active nucleons
\begin{equation}
\frac{1}{2}A(A-1)\frac{(2\pi)^3}{m}\rho(\bl_1,\bl_2|\bl'_1,\bl'_2).
\label{0c}
\end{equation}

Next step is integration over the transverse momenta of the active nucleons
$l_{1\perp},l_{2\perp},l'_{1\perp}$ and $l'_{2\perp}$ with
\begin{equation}
l_{1\perp}+l_{2\perp}=l'_{1\perp}+l'_{2\perp}.
\end{equation}
The high-energy part is independent of these transverse momenta.
We present
\begin{eqnarray}
2\pi\delta^2(l_{1\perp}+l_{2\perp}-l'_{1\perp}-l'_{2\perp})=
\int \frac{d^2b}{2\pi}e^{ib(l_{1\perp}+l_{2\perp}-l'_{1\perp}-l'_{2\perp})}
\end{eqnarray}
and pass  to the transverse coordinate space in the density matrix.
We get (suppressing the dependence on $z$-components)
\begin{equation}
\frac{1}{(2\pi)^9}\int \prod_{i=1,2}d^2l_id^2l'_i
\prod_{i=1,2}d^2x_id^2x'_id^2b
e^{ib(l_{1\perp}+l_{2\perp}-l'_{1\perp}-l'_{2\perp})}\]\[
e^{-ix_1l_{1\perp}-ix_2l_{2\perp}+ix'_1l'_{1\perp}+ix'_2l'_{2\perp}}
\rho(x_1,x_2|x'_1,x'_2)=\int\frac{d^2b}{2\pi}\rho(b,b|b,b).
\end{equation}

We are left with integrations over the $z$-components $l_{1z},l_{2z}$
and $\lambda_z$ with the high-energy part depending only on $\lambda_z$.
So our  expression for the amplitude is
\begin{equation}
{\cal A}=A(A-1)\frac{(2\pi)^3}{2m}
\int\frac{d^2bdl_{1z}dl_{2z}d\lambda_z}{(2\pi)^4}H(\lambda_{z})
\rho(b,l_{1z};b,l_{2z}|b,l_{1z}+\lambda_z;b,l_{2z}-\lambda_z).
\end{equation}
Again we pass to the coordinate representation for the density matrix
as a function of $z$-components of momenta to obtain
\begin{eqnarray}
{\cal A}=
\frac{A(A-1)}{4\pi m}
\int d^2bd\lambda_zH(\lambda_{z})dz_1dz_2
e^{i\lambda_z(z_1-z_2)}
\rho(b,z_1;b,z_2|b,z_1;b,z_2).
\label{0b}
\end{eqnarray}
This is our final expression. Integration over $z_1-z_2$ is carried out
along the path passing through the nucleus of the average length
$\sim R_A\sim A^{1/3}$ and large in the limit $A>>1$. Correspondingly
the order of essential values of $\lambda_z$ is $\sim A^{-1/3}$.
Taking the imaginary part we obtain the inclusive cross-section as
\beq
\frac{(2\pi)^3d\sigma}{d^3p}=\frac{1}{s}{\rm Im}\,{\cal A}=
\frac{A(A-1)}{4\pi ms}
\int d^2bd\lambda_z{\rm Im}\,H(\lambda_{z})dz_1dz_2
e^{i\lambda_z(z_1-z_2)}
\rho(b,z_1;b,z_2|b,z_1;b,z_2).
\label{inclugen}
\eeq

In fact the high-energy part normally contains a pole
at $\lambda_z=0$ ~\cite{gribov,CK},
so that its imaginary part can be presented as
\beq
{\rm Im}\,H(\lambda_z)=F\delta(\lambda_z)+C(\lambda_z).
\eeq
The term with $\delta(\lambda_z)$  gives the standard Glauber contribution corresponding to
multiple collisions of the incident particle on the nucleons located
at large distances between one another.
Integration over $\lambda_z$  gives :
\beq
\frac{(2\pi)^3d\sigma}{d^3p}=\frac{A(A-1)}{4\pi ms}F
\int d^2bdz_1dz_2
\rho(b,z_1;b,z_2|b,z_1;b,z_2).
\label{gla1}
\eeq
If we neglect nuclear correlations and
factorize the density matrix (\ref{gla1}) transforms into the
standard Glauber expression
\beq
\frac{(2\pi)^3d\sigma}{d^3p}=
\frac{A(A-1)}{4\pi ms}F\int d^2b T^2(b),
\eeq
where $T_A(b)$ is the standard nuclear profile function
\beq
T(b)=\int dz\rho(b,z).
\eeq
For $A>>1$ this Glauber contribution has order $A^{4/3}$.
The non-singular part $C(\lambda_z)$ gives a contribution
corresponding to the scattering of the incident particle on two nucleons
located at the same point. Indeed at $\lambda_z\to 0$ one can take
$C(\lambda_z)$ out of the integration  over $\lambda_z$ in (\ref{0b}),
so that this integration gives $2\pi\delta(z_1-z_2)$.
At $A>>1$ the resulting contribution has order $A$ and
and can be neglected as compared to (\ref{gla1}).
So the leading Glauber term is determined
by the contribution to $H(\lambda_z)$ singular in the limit
$\lambda_z\to 0$.

In fact we shall find out that the situation is more complicated.
Many of our contributions to Im~$H(\lambda_z)$ will contain terms
proportional to $\delta(\lambda_z-\beta(q_{\perp}))$ under the sign
of integration over transverse momentum $q_{\perp}$ of the
exchanged reggeons.
In the limit when the energies of both the projectile and observed gluon
tend to infinity all $\beta$ tend to zero, so that one finally obtains
a non-zero contribution to the Glauber cross-section. However one finds
that $\beta$ diminishes either with the energy of the projectile or
with the energy of the observed gluon, much smaller than the former:
\[\beta(q_\perp)\sim \frac{mq_\perp^2}{s}\ \ {\rm or}\ \
\beta(q_\perp)\sim \frac{mq_\perp^2}{xs}\]
with $x=p_+/k_+<<1$.
Put in (\ref{inclugen}) this introduces an oscillating factor into
the integrand
\[e^{(z_1-z_2)\frac{mq_\perp^2}{s}}
\ \ {\rm or} \ \ e^{(z_1-z_2)\frac{mq_\perp^2}{sx}}.\]
At very large $s$ this factor turns to unity. This happens when
\beq
R_A\frac{mq_\perp^2}{xs}<<1.
\label{kincon}
\eeq
In the following we shall assume that this condition is fulfilled. Then
we can neglect all terms depending on the transverse momenta as compared to
$\lambda$ in the integrands. In particular we can neglect
$p_-=-p_\perp^2/2p_+$ where it enters with $\lambda_-=-\lambda_z/\sqrt{2}$
and integration variables.

Function $H(\lambda_z)=H(p\lambda,k\lambda)$ can be calculated in any
system and in
the c.m. system $k_z+l_z=k_{\perp}=l_{\perp}=0$ in particular,
 using the fact that the
relevant scalar products are Lorenz invariant. For the calculation it is
instructive to see the relative orders of all scalar product on
which $H$ depends. They are $(kp), (kl), (pl), (k\lambda)$ and $(p\lambda)$.
We define the longitudinal momenta as $a_{\pm}=a_0\pm a_z$.
In the lab. system, denoting the $\pm$ momenta with tildes,
 we have $\tilde{l}_{\pm}=m$,
$\tilde{\lambda}_{\pm}=\mp \lambda_z$, so that $\lambda_-
=-\lambda_z(\sqrt{s}/m)$
The total energy squared is $s=2(kl)=k_+l_-=\tilde{k}_+ m$, from which
we find $\tilde{k}_+=s/m$ and then $\tilde{p}_+=xs/m$. We also have
$\tilde{k}_-=0$ and $\tilde{p}_--mp_\perp^2/xs$.
Furthermore
\begin{equation}s_1=(k+p)^2=2(kp)=k_+p_-=-\frac{p_\perp^2}{x},\ \
 s_2=(p+l)^2=p_+l_-=xs,
\end{equation}
(as expected $s_1s_2=-p_\perp^2 s$). We assume all the three $s,s_1$ and
$s_2$ large, which requires $x<<1$.
Using the order of $\lambda_z$ in the lab system
we find
$\lambda_\perp=0$ so that
$
kq=\sim s \sqrt{\frac{\epsilon}{m}}
$
and
$
pq=\sim xs \sqrt{\frac{\epsilon}{m}}
$
So we have relative orders
\begin{eqnarray}
\frac{k\lambda}{kl}\sim\sqrt{\frac{\epsilon}{m}},\ \
\frac{k\lambda}{pl}\sim\frac{1}{x}\sqrt{\frac{\epsilon}{m}},\ \
\frac{k\lambda}{k\lambda}\sim \frac{sx}{p_\perp^2}\sqrt{\frac{\epsilon}{m}},\ \
\frac{p\lambda}{kl}\sim x\sqrt{\frac{\epsilon}{m}},\ \
\frac{p\lambda}{pl}\sim\sqrt{\frac{\epsilon}{m}},\ \
\frac{p\lambda}{kp}\sim \frac{s}{p_\perp^2}\sqrt{\frac{\epsilon}{m}}.
\end{eqnarray}
In all cases the limit $\lambda_z\to 0$ is obtained taking $\epsilon/m \to 0$
and therefore neglecting $(k\lambda)$ and $(p\lambda)$ as compared to
$(kl),(pl)$ and $(kp)$.

This fixes the rules for calculating the high-energy part $H$.
Assuming for simplicity that the nucleus consists of quarks,
one can take the two initial target quark momenta equal to $l$,
the two final momenta of the target quark as $ l+\lambda$
and $ l-\lambda$ with $\lambda_\perp=\lambda_+=0$.
Under assumption (\ref{kincon}) one can  neglect all terms
depending on the transverse momenta ( $p_-$ in particular)
as compared to $\lambda_-$ and ``-''-components of integration momenta.
Since these terms contain
$k_+$ or $p_+$ in the denominator, this also corresponds to taking the
limit $k_+,p_+\to\infty$ in the final formulas. Then one has to take
the limit $\lambda\to 0$ leaving only terms singular in this limit.

\section{Non-diffractive contribution to the inclusive gluon production from
R$\to$R(R)P vertices: the diagrams}

\begin{figure}[h]
\leavevmode \centering{\epsfysize=0.3\textheight\epsfbox{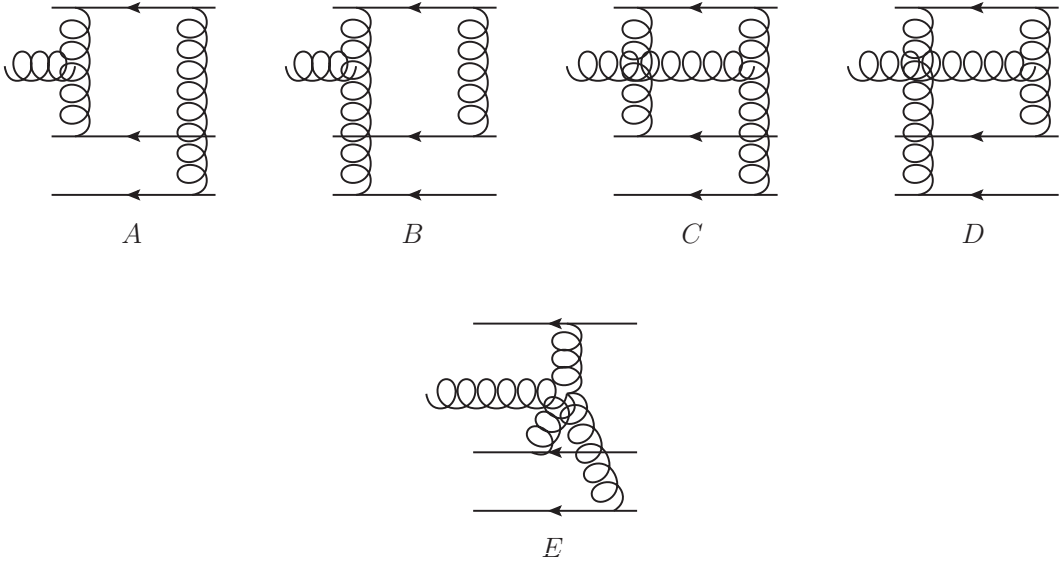}}
\caption{Gluon production amplitudes off two targets}
\label{fig2}
\end{figure}

 In the framework of the effective action \cite{lipatov},
in the lowest order
the non-diffractive inclusive cross-section on two scattering centers
("nucleons") generated by
the R$\to $R(R)P vertices is given by the square modulus of the sum
of five production amplitudes, shown in Fig.~\ref{fig2} and denoted as
A,B,C,D and E. In the first four the number of reggeons does not change
and the gluon is emitted by the R$\to$RP vertex. In Fig.~\ref{fig2},E the
number of reggeons changes and the gluon is emitted by the R$\to$RRP vertex.
In the kinematical
region $q_{1-},q_{2-}\sim \lambda_->>p_-$
essential for the inclusive cross-section,  the diagrams A$-$D
give only contributions to the production amplitude proportional to
$\delta(q_{1-})$ or $\delta(q_{2-})$
and the diagram E contributes the rest ~\cite{new}. However, alternatively
one can include diagrams A$-$D as a whole but then one has to exclude
a part of the contribution from diagram E containing the principal part
of the pole contribution at $q_{1,2-}=0$  ~\cite{new}. We shall use
the latter approach and discuss this point in the next section.

Taking the square modulus of the sum of these production amplitudes
gives rise to 25 scattering amplitudes cut in the center, which are shown
in Figs.~\ref{fig3}--\ref{fig7}. For interference diagrams only half is
shown, the other half given by their complex conjugates.
Note that the total contribution of these amplitudes to the cross-section
includes also discontinuities across  other cuts passing through only
one of both targets. They involve the R$\to$RRRP vertex and,
as mentioned in the Introduction, will be considered in a separate
publication.

As mentioned, for simplicity
we model the two nuclear components
by a pair of quarks or a pair of quark and antiquark with momenta
$l_{1}$ and $l_{2}$ over which one has to integrate with the
nuclear wave functions.
Both $qq$ and $q\bar{q}$ targets lead to the same result.
Here we will consider factors general for all these diagrams.

We choose $q_{1}$, $q_{1}^{'}$,
$l_{1}$ and $l_{1}^{'}$ as integration variables.
In the following we define (see Figs. \ref{fig3}--\ref{fig7}
for notations)
\begin{equation}
\lambda=l_{1}^{'}-l_{1}=l_2-l_{2}^{'},\ \
q_{1}^{'}=q_{1}-\lambda,\ \
q_{2}^{'}=q_{2}+\lambda,\ \
q=p+q_{1}+q_{2}
\label{1}
\end{equation}
with $\lambda_+=\lambda_\perp=0$.

We work in the Regge kinematics, so that the longitudinal
components of our momenta should obey:
\begin{equation}
k_{+}\gg p_{+}\sim q_{+}\gg l_{1+}
\sim l_{2+}\sim l'_{1+}\sim l'_{2+}\sim q_{1+}
\sim q_{2+}\sim q'_{1+}\sim q'_{2+},
\label{2}
\end{equation}
\begin{equation}
l_{1-}\sim l'_{-}\sim l_{2-}\sim l'_{2-}
\gg p_{-}\sim q_{1-}\sim q_{2-}\sim q'_{1-}\sim q'_{2-}\gg k_{-}\sim q_{-}.
\label{3}
\end{equation}
The quark masses are assumed to be equal to zero and we can put
$k_-=k_{\perp}=l_{i+}=l_{i\perp}=0$ for $i=1,2$ Each
transversal momentum is assumed to be much smaller than the larger
of the longitudinal one.

Factors corresponding to the two target quarks, projected
onto colourless states,  are
\beq
2\pi \frac{g^2}{2N_c}l_{1-}^2\delta((l_{1}+q_{1})^{2})\ \
{\rm and}\ \
2\pi \frac{g^2}{2N_c}l_{2-}^2\delta((l_{2}+q_{2})^{2})
\label{target}
\eeq
We are interested in terms singular in $\lambda_-$.

The total number of longitudinal
integrations is, obviously, four. However three of them are removed
by the mass-shell conditions for real particles:
\begin{equation}
(k-q)_-=-\frac{q_\perp^2}{(k-q)_+},\ \
(l_1+q_1)_+=-\frac{q_{1\perp}^2}{(l_1+q_1)_-},
\ \
(l_2+q_2)_+=-\frac{q_{2\perp}^2}{(l_2+q_2)_-}.
\label{28}
\end{equation}
The  last two conditions fix $q_{1+}$ and $q_{2+}$. From the first we
find $q_-$ and from this relate
$
q_{1-}+q_{2-}=(q-p)_-\simeq -p_-
$
We also have $q'_1=l_1+q_1-l'_1$ and $q'_2=l_2+q_2-l'_2$ and the
emitted gluon momenta are (from the right) $q-q_1$, $q-q_1-q_2=p$ and
$q-q_1-q_2+q'_1$. Thus we are left with a single longitudinal variable
for which we choose $q_{1-}$.
So in our formulas  momentum integrations are:
\begin{equation}
\frac{4}{(2\pi)^8 k_+l_-^2}\int d^2q_{1\perp}d^2q_{2\perp}dq_{1-}
\frac{1}{q_{1\perp}^4q_{2\perp}^4},
\label{29}
\end{equation}
where we used that in our kinematics $l_{1-}=l_{2-}=l_{-}/2=\sqrt{s}/2$,
$q'_{1\perp}=q_{1\perp}$ and $q'_{2\perp}=q_{2\perp}$.

\begin{figure}[h]
\leavevmode \centering{\epsfysize=0.2\textheight\epsfbox{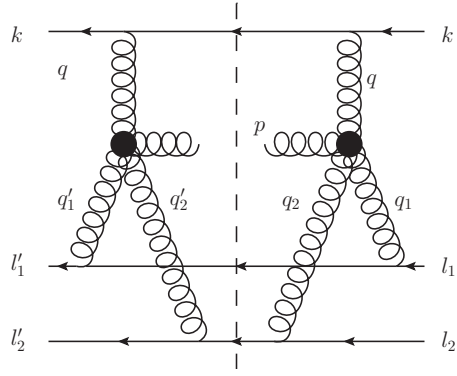}}
\caption{Contribution  from two R$\to$RRP vertices}
\label{fig3}
\end{figure}

\section{Contribution  from the R$\to$RRP vertex (diagram Fig. \ref{fig3})}

We start with the diagram in Fig. \ref{fig3}, which  comes from the
product of production amplitudes $(E)\cdot (E)^*$
The product of reggeon propagators
(apart from those included into the weight) in our
approximation reduces to $1/q^4_\perp$.
It depends only on the transversal components.
The factor corresponding to the projectile quark $k$ is:
\begin{equation}
2\pi g^{2}\delta((k-q)^2)k_{+}^{2}.
\label{5}
\end{equation}

The R$\to$RRP vertex $q\to q_{1}+q_{2}+p$ was derived in \cite{bravyaz}.
It consists of two parts:
\begin{equation}
V=V_1+V_2,
\label{11}
\end{equation}
where
\begin{equation}
V_1=ig^2\frac{f^{ab_1d}f^{b_2cd}}{(q-q_1)^2+i0}
\Big\{q_+(q,\ep)-\frac{q_\perp^2}{q_{1-}}
\Big[(q-q_1,\ep)-\frac{(q-q_1)^2}{p_\perp^2}(p,\ep)\Big]\Big\}
\label{12}
\end{equation}
and
\begin{equation}
V_2=
ig^2\frac{f^{ab_2d}f^{b_1cd}}{(q-q_2)^2+i0}
\Big\{q_+(q,\ep)-\frac{q_\perp^2}{q_{2-}}
\Big[(q-q_2,\ep)-\frac{(q-q_2)^2}{p_\perp^2}(p,\ep)\Big]\Big\}.
\label{13}
\end{equation}

\begin{figure}[h]
\leavevmode \centering{\epsfysize=0.12\textheight\epsfbox{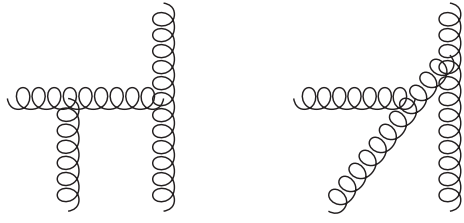}}
\caption{Two parts $W$ and $R$ of the R$\to$RRP vertex}
\label{fig4}
\end{figure}

It is convenient to split the terms in each of the parts $V_1$ and $V_2$
into two terms depending on their singularities in $q_{1-}$ and $q_{2-}$
respectively. Take $V_1$. In the second term in the brackets
we present the product $1/q_{1-}[(q-q_1)^2+i0]$ as
\[
\frac{1}{q_{1-}[(q-q_1)^2+i0]}=
-\frac{1}{q_{+} a}\Big(\frac{1}{q_{1-}-a-i0}-\frac{1}{q_{1-}}\Big)
=\frac{1}{a[(q-q_1)^2+i0]}
+\frac{1}{q_{1-}(q-q_1)_\perp^2},
\]
where
\beq
a=(q-q_1)_\perp^2/q_+.
\label{a}
\eeq
Combining terms with the same singularity in $q_{1-}$
we find
\beq
V_1=W_1+R_1,
\eeq
where
\beq
W_{1}=ig^2\frac{q_+q_{\perp}^2}{(q-q_1)^2+i0}
\bbb(p,q_2,q_1)\be_\perp f^{ab_1d}f^{b_2cd}
\eeq
and
\beq
R_{1}=-ig^2\frac{q_{\perp}^2}{q_{1-}}
\bbl(p,q_2)\be_\perp f^{ab_1d}f^{b_2cd}.
\eeq
Here the transverse spatial vectors $\bbb$ and $\bbl$ are the Bartels and
Lipatov vertices respectively:
\beq
\bbl(p,q_2)=\frac{\bp}{p_\perp^2}-\frac{\bp+\bq_2}{(p+q_{2})_\perp^2},\ \ \
\bbb(p,q_2,q_1)=\bbl(p+q_2,q_1)
\label{effvert}
\eeq
and we use $q-q_1=p+q_2$. Production amplitudes corresponding
to the two terms $W_1$ and $R_1$ are schematically
illustrated in Fig. \ref{fig4}.
A similar representation for $V_2$ is obtained by interchanging the
target quarks.

As mentioned, the  study of the production amplitude in the framework of the
effective action and in the kinematics $q_{1,2-}>>p_-$
demonstrates that to get its correct parts, symmetric and
antisymmetric with
respect to the change $q_{1}\lra q_{2}$, one should
drop the contribution of $R_{1,2}$ but include diagrams A,$...$,D
~\cite{new}.
So considering the contribution from the R$\to$RRP vertex we restrict
ourselves to the contribution from from $W_{1,2}$ only.

Taking the square modulus, this contribution is a sum
four parts, symbolically
\begin{equation}
(W_{1})\cdot (W_{1})^*+
(W_{2})\cdot (W_{2})^*+
(W_{1})\cdot (W_{2})^*+
(W_{2})\cdot (W_{1})^*.
\label{17}
\end{equation}
Obviously, the second term is obtained from the first one and
the fourth term from the third one by the interchange
$\lambda\to -\lambda$.
So it is sufficient to calculate $(W_1)\cdot (W_1)^{*}$ and
$(W_{1})\cdot (W_{2})^{*}$.

Different parts carry different color factors.
Combining with (27) the total
color factor for the part
$(W_{1})\cdot (W_{1})^{*}$ is
\begin{equation}
C_{11}
=\frac{N^2_c-1}{8N_c}.
\label{18}
\end{equation}
For the part
$(W_{1})\cdot (W_{2}^{*})$ it is twice smaller:
\begin{equation}
C_{12}=\frac{N_c^2-1}{16N_c}.
\label{21}
\end{equation}

{\bf 1.} $(W_1)\cdot (W_1)^*$

Neglecting the dependence on
$q_{1-}$ of the target, the remaining dependence comes from the
denominators of the two virtual gluon propagators
\begin{equation}
D_{1}=(q-q_1)^2+i0
\label{31}
\end{equation}
and
\begin{equation}
D_{2}=(q-q'_1)^2-i0=(q-q_1+\lambda)^2-i0 .
\label{33}
\end{equation}
The integral over the longitudinal variable is
\begin{equation}
I_{11}=\int dq_{1-}\frac{1}{((q-q_1)^2+i0)
((q-q_1+\lambda)^2-i0)}
=I(-q,-q-\lambda) .
\label{int1n}
\end{equation}
Here for future use we define
\[
I(p_1,p_2)=\int dq_{1-}
\frac{1}{((q_1 + p_1)^2 +i0)((q_1 + p_2)^2 -i0)}=
\]
\begin{equation}
=\frac{1}{p_{1+}p_{2+}} \int dq_{1-}
\frac{1}{ \Big( q_{1-}+p_{1-}+\frac{(q_1 + p_1)_{\perp}^2}{p_{1+}}
 +i0\cdot p_{1+} \Big)
\Big( q_{1-}+p_{2-}+\frac{(q_1 + p_2)_{\perp}^2}{p_{2+}}
 -i0\cdot p_{2+} \Big) }\ .
\label{int0}
\end{equation}
When $p_{1+}p_{2+}>0$ in (\ref{int0}),
the two poles in $q_{1-}$ lie on the opposite
sides of the real axis. Taking the residue we find
\begin{equation}
I(p_1,p_2)=
\frac{2\pi i}{p_{1+}p_{2+}}\cdot
\frac{-{\rm sign}(p_{1+})}{p_{2-}-p_{1-}+\frac{(q_1 + p_2)_{\perp}^2}{p_{2+}}
-\frac{(q_1 + p_1)_{\perp}^2}{p_{1+}}-i0\cdot {\rm sign}(p_{1+})}\ .
\label{int01}
\end{equation}
When $p_{1+}p_{2+}<0$, the two poles in $q_{1-}$ lie on the same
side of the real axis and the integral equals to zero.

For (\ref{int1n}) we use $q'_{1\perp}=q_{1\perp}$,
so $(q-q_1)_{\perp}^2=(q-q'_1)_{\perp}^2$,
and also assume (as always in the Regge kinematics) $q_{-}<< q_{1-}$,
then
\begin{equation}
I_{11}=I(-q,-q-\lambda)
=\frac{2\pi i}{q_+^2}\frac{1}{-\lambda_{-}+i0}.
\label{int1}
\end{equation}
The result has a pole at $\lambda_1=0$ and will give a non-zero
contribution to the inclusive cross-section.

{\bf 2.} $(W_1)\cdot (W_2)^*$

The difference  from the term $(W_1)(W_1)^*$ is
in $D_2$, which is now changed to
\begin{equation}
D_2=(p+q'_1)^2-i0=(p+q_1-\lambda)^2-i0 .
\end{equation}
We get an integral
\begin{equation}
I_{12}=I(-q,p-\lambda)=0 ,
\end{equation}
which is equal to zero
since both poles are located in the upper half-plane.

{\bf 3.}

The total contribution from $W_{1,2}$ is given by the sum
$(W_{1})\cdot (W_{1})^{*}+(\lambda\to -\lambda)$.
We find
\beq
(W_{1})\cdot (W_{1})^{*}+(\lambda\to -\lambda)=
4\pi^2\delta(\lambda_-)q_\perp^4
\bbb^2(p,q_2,q_1).
\eeq
Supplying all the rest factors and using the relation between $\lambda_-$
in the c.m. system
and $\lambda_z$ in the lab. system we get for the factor $F$
in the high-energy part coming from the product $|(W)_1+(W)_2|^2$
\begin{equation}
F_1=
4\pi msg^{10}\frac{N_c^2-1}{N_c}
\int\frac{d^2q_{1\perp}d^2q_{2\perp}}{(2\pi)^4}
\frac{1}{q^4_{1\perp}q^4_{2\perp}}
\bbb^2(p,q_2,q_1).
\label{50ee}
\end{equation}
This is the same expression which follows in the purely transversal
BFKL-Bartels approach.

\section{Contributions  from the R$\to$RP vertex}

This contribution comes from the diagrams in which
the number of reggeons does not change and the gluons are emitted
by the Lipatov verices.
It is generated by the square modulus of the sum
of production amplitudes $(A)+(B)+(C)+(D)$ in Fig. \ref{fig2}).
It can be split into a direct part $\sum$(I)$\cdot$(I)$^*$, $I=A,B,C,D$
and the interference part  $\sum$(I)$\cdot$(J)$^*$, $I\neq J=A,B,C,D$.
The factor corresponding to the projectile quark $k$ is $2\pi g^{4}\delta((k-q)^2)k_{+}^{4}$.

\subsection{The direct part}

\begin{figure}[h]
\leavevmode \centering{\epsfysize=0.2\textheight\epsfbox{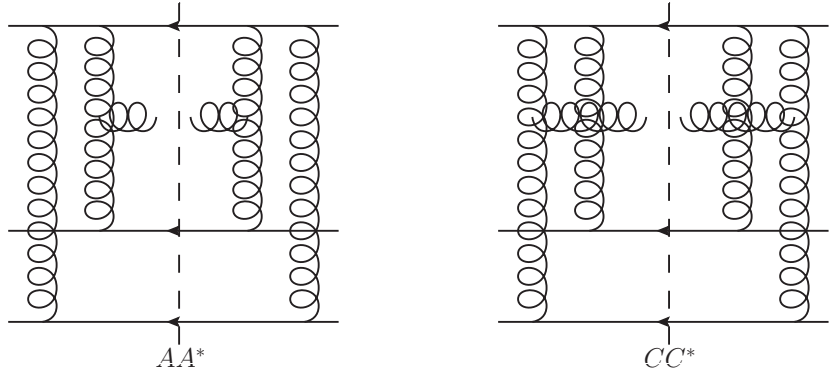}}
\caption{Direct diagrams with two R$\to$RP vertices}
\label{fig5}
\end{figure}

Diagrams corresponding to products $(A)\cdot(A)^*$ and $(C)\cdot(C)^*$
are shown in Fig. \ref{fig5}. The other two corresponding to
products  $(B)\cdot(B)^*$ and $(D)\cdot(D)^*$ are obtained by the interchange
of target quarks 1 and 2, which reduces to the change $\lambda\to -\lambda$.
So they need not be studied separately.

{\bf 1.} $(A)\cdot (A)^*$

The product of reggeon propagators and R$\to$RP vertices,
summed over polarizations of the observed gluon, gives
\beq
g^2f^{a_1b_1c}f^{a'_1b'_1c}\bbl(p,q_1)\bbl(p,q'_1),
\label{prveaa}
\eeq
where vector $\bbl$ is  the Lipatov vertex (\ref{effvert}).

The colour factor is found to be
\begin{equation}
C_{AA}=\frac{(N_c^2-1)^2}{16N_c^3}.
\label{155}
\end{equation}

The dependence on $q_{1-}$ comes from the virtual quarks propagators
with the denominators
\begin{equation}
D_{1}=(k-q_2)^2+i0=(k+p-q+q_1)^2+i0
\label{157}
\end{equation}
and
\begin{equation}
D_{2}=(k-q_{2}')^2-i0=(k+p-q+q_1-\lambda)^2 -i0 .
\label{158}
\end{equation}
Integration over $q_{1-}$ gives
\begin{equation}
I_{AA}=I(k+p-q,k+p-q-\lambda)
=\frac{2\pi i}{k_{+}^2}\frac{1}{\lambda_{-}+i0} ,
\label{159}
\end{equation}
where the relation $q'_{2\perp}=q_{2\perp}$ is used.

{\bf 2.} $(C)\cdot (C)^*$

The reggeon propagators and vertices are obtained from (\ref{prveaa})
by the change $q_1,b_1\to q_2,b_2$ and $q'_1,b'_1\to q'_2,b'_2$.
The dependence on $q_{1-}$ comes from the denominators in the
virtual quarks propagators:
\begin{equation}
D_{1}=(k-p-q_2)^2+i0=(k-q+q_1)^2+i0,
\label{194}
\end{equation}
\begin{equation}
D_{2}=(k-p-q_2')^2-i0=(k-q+q_1-\lambda)^2-i0.
\label{195}
\end{equation}
We get the longitudinal integral
\begin{equation}
I_{CC}=I(k-q,k-q-\lambda)
=\frac{2\pi i}{k_+^2}\frac{1}{\lambda_{-}+i0}.
\label{196}
\end{equation}

{\bf 3.}

Taking into account that the contributions from $(B)\cdot(B)^*$
and $(D)\cdot(D)^*$ are obtained from those from
$(A)\cdot(A)^*$ and $(C)\cdot(C)^*$ respectively by the change
$\lambda\to -\lambda$ we find
\begin{equation}
(A)\cdot (A)^*+(B)\cdot (B)^*=
8\pi g^{10}l_{1-}l_{2-}k_{+}\frac{(N_c^2-1)^2}{N_c^3}\delta(\lambda_-)
\int\frac{d^2q_{1\perp}d^2q_{2\perp}}{(2\pi)^4}
\frac{1}{q^4_{1\perp}q^4_{2\perp}}\bbl^2(p,q_1).
\label{206}
\end{equation}
The contribution from $(C)\cdot(C)^*+(D)\cdot (D)^*$ is obtained from
(\ref{206}) by interchanging $q_{1\perp}\leftrightarrow q_{2\perp}$.
So we finally get for the
contribution to the high-energy part
\begin{equation}
F_2=
4\pi msg^{10}\frac{(N_c^2-1)^2}{N_c^3}
\int\frac{d^2q_{1\perp}d^2q_{2\perp}}{(2\pi)^4}
\frac{1}{q^4_{1\perp}q^4_{2\perp}}\bbl^2(p,q_1).
\label{208}
\end{equation}
In the integrand we find the product of two Lipatov vertices.
The same result is obtained in the BFKL-Bartels approach.

\subsection{Interference contributions from $|(A)+(B)+(C)+(D)|^2$}

\begin{figure}[h]
\leavevmode \centering{\epsfysize=0.35\textheight\epsfbox{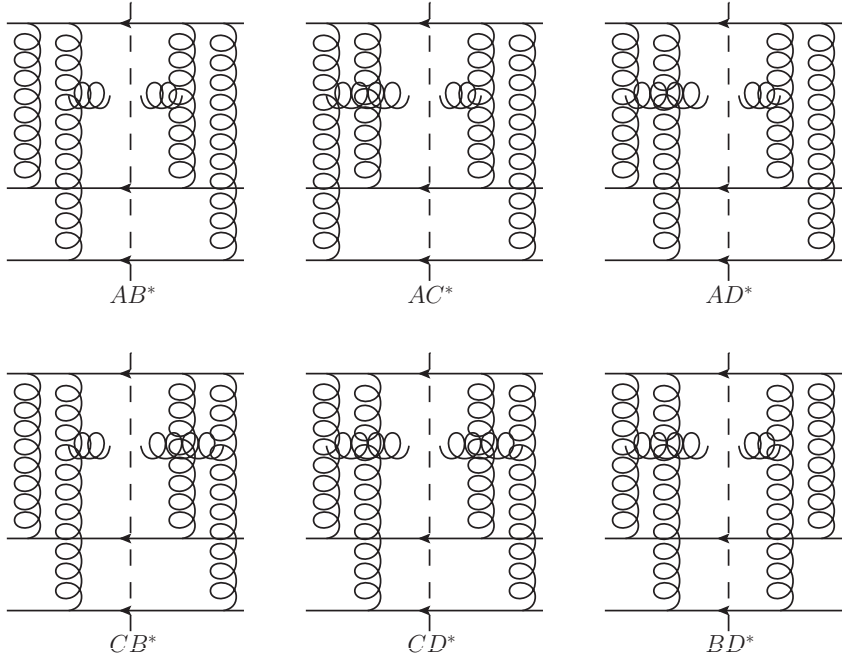}}
\caption{Interference diagrams with two R$\to$RP vertices}
\label{fig6}
\end{figure}

The 12 interference contributions from $|(A)+(B)+(C)+(D)|^2$
is the sum
\[(A)\cdot (B)^*+(A)\cdot (C)^*+(A)\cdot (D)^*
+(C)\cdot (B)^*+(C)\cdot (D)^*+(B)\cdot (D)^*,\]
the diagrams for which are shown in Fig. \ref{fig6}
and its conjugates.
Obviously, $(B)\cdot (D)^*$ is obtained from $(A)\cdot (C)^*$
by the interchange of target quarks, that is, the change
$\lambda\to -\lambda$.  The same interchange transforms $(A)\cdot (D)^*$
into $(B)\cdot (C)^*=[(C)\cdot (B)^*]^*$, which amounts to changing
$\lambda\to-\lambda$ and taking conjugate. So it is sufficient
to consider 4 diagrams $(A)\cdot (B)^*$, $(A)\cdot (C)^*$,
$(A)\cdot (D)^*$ and $(C)\cdot (D)^*$.

{\bf 1.} $(A)\cdot (B)^*$

The product of reggeon propagators and $R\to RP$ vertices is
\beq
g^2f^{a_1b_1c}f^{a'_1b'_2c}\bbl(p,q_1)\bbl(p,q'_2).
\eeq
The colour factor is
\begin{equation}
C_{AB}=0.
\label{165}
\end{equation}
The dependence on $q_{1-}$ comes from the denominators in the
virtual quark propagators, which are $D_1$ from
(\ref{157}) and
\begin{equation}
D_{2}=(k-q'_1)^2-i0=(k-q_1+\lambda)^2-i0 .
\label{167}
\end{equation}
We get the longitudinal integral
\begin{equation}
I_{AB}=I(k+p-q,-k-\lambda)=0 .
\label{168}
\end{equation}

{\bf 2.} $(A)\cdot (C)^*$

The product of reggeon propagators and R$\to$RP vertices is the same
as for $(A)(B)^*$.
The colour factor is
\begin{equation}
C_{AC}=-\frac{N_c^2-1}{32N_c}.
\label{172}
\end{equation}
The dependence on $q_{1-}$ comes from the denominators $D_1$
from (\ref{157}) and
\begin{equation}
D_{2}=(k-p-q_{2}')^2-i0=(k-q+q_{1}-\lambda)^2-i0 .
\label{174}
\end{equation}
We get the integral
\begin{equation}
I_{AC}=I(k+p-q,k-q-\lambda)
=\frac{2\pi i}{k_+^2}
\frac{1}{\lambda_{-}+\frac{q_{2\perp}^2}{k_{+}}
-\frac{(p+q_2)^2_{\perp}}{k_{+}}+i0},
\label{175}
\end{equation}
where we neglected the term $p_{-}$ as compared to $\lambda_{-}$
in the denominator.
In the limit $k_+\to\infty$ this develops a pole at $\lambda_-=0$
and will give a non-zero contribution to the inclusive cross-section.

{\bf 3.} $(A)\cdot (D)^*$

The product of reggeon propagators and $R\to RP$ vertices
is the same as for the term $(A)\cdot (A)^*$ (\ref{prveaa}).
The colour factor is
\begin{equation}
C_{AD}=-\frac{N_c^2-1}{16N_c^3}.
\label{178}
\end{equation}
The denominators are $D_1$ from (\ref{157}) and
\begin{equation}
D_{2}=(k-q+q_{2}')^2-i0=(k-p-q_1+\lambda)^2-i0 .
\label{180}
\end{equation}
We get the integral
\begin{equation}
I_{AD}=I(k+p-q,-k+p-\lambda)=0 .
\label{181}
\end{equation}

{\bf 4.} $(C)\cdot (D)^*$

The product of reggeon propagators
R$\to$RP vertices is the same as for $(C)\cdot (C)^*$
The color factor is given by (\ref{165}) and equal to zero.
The dependence on $q_{1-}$ comes from the denominators
$D_1$ from (\ref{194}) and $D_2$ from (\ref{180}).
We get the integral
\begin{equation}
I_{CD}=I(k-q,-k+p-\lambda)=0 .
\label{199}
\end{equation}

{\bf 5.}

As a result,
interference contribution comes from diagrams
\[\Big[(A)\cdot (C)^*+(\lambda\to -\lambda)\Big]+c.c.\]
Supplying all the rest factors we find
\beq
F_3=-2\pi msg^{10}\frac{N_c^2-1}{N_c}\int\frac{d^2q_{1\perp}d^2q_{2\perp}}
{(2\pi)^4}\frac{1}{q_{1\perp}^4q_{2\perp}^4}\bbl(p,q_1)\bbl(p,q_2).
\eeq

In the integrand there naturally appears product of Lipatov vertices.
The same result follows from the BFKL-Bartels approach.

\section{Interference between the R$\to$RP and R$\to$ RRP vertices}

\begin{figure}[h]
\leavevmode \centering{\epsfysize=0.45\textheight\epsfbox{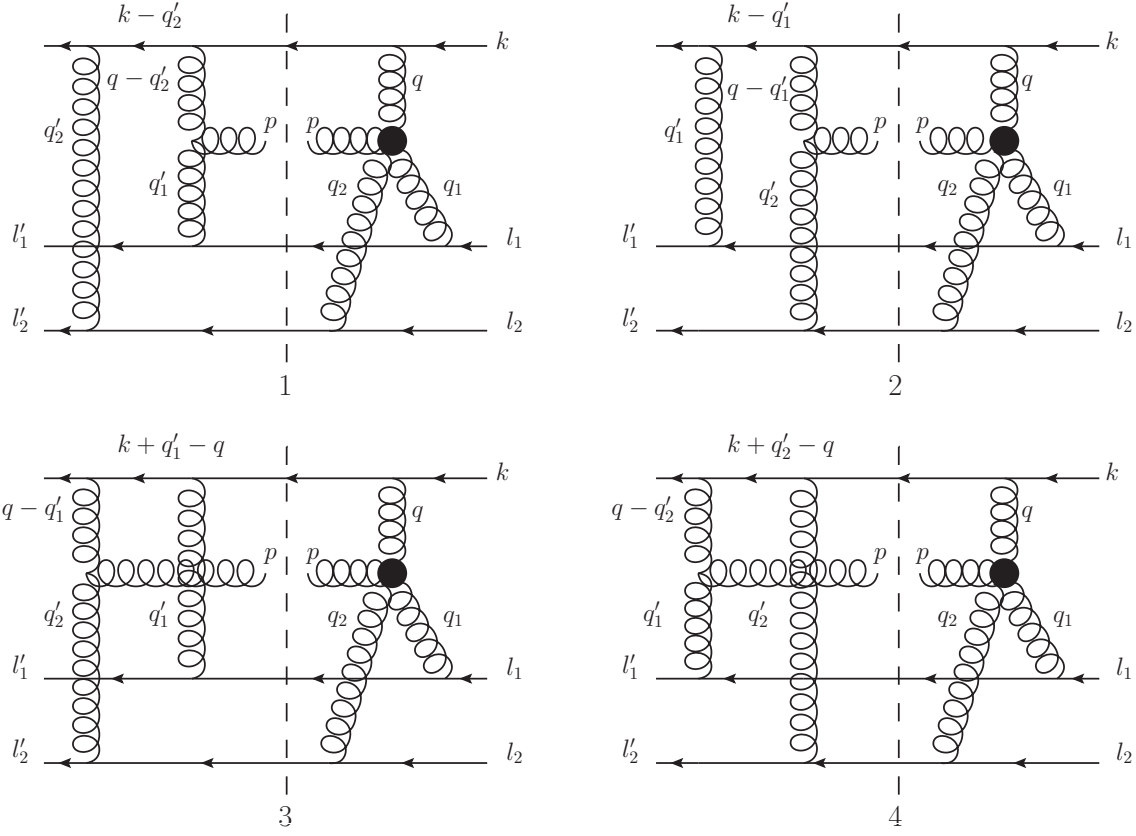}}
\caption{Interference between the R$\to$RP and R$\to$ RRP vertices}
\label{fig7}
\end{figure}

It corresponds to 4 diagrams $(E)\cdot [(A)^*+(B)^*+(C)^*+(D)^*]$
shown in in Fig. \ref{fig7}
plus 4 conjugated diagrams $[(A)+(B)+(C)+(D)]\cdot (E)^*$
We consider only diagrams shown in Fig. \ref{fig7}. Obviously,
$(E)\cdot (B)^*$ and $(E)\cdot (D)^*$ are obtained from $(E)\cdot (A)^*$
and $(E)\cdot (C)^*$ respectively by the interchange of the target quarks
which is realized by the change $\lambda\to -\lambda$. So it is sufficient
to consider only two diagrams $(E)\cdot (A)^*$ and $(E)\cdot (C)^*$.

Factors corresponding to the target quarks and the
R$\to$ RRP vertex $V$
are common to all the diagrams and can be borrowed from
(\ref{target}), (\ref{12}) and (\ref{13}).
The spin-momentum part of the factor corresponding to the projectile quark,
also common for all the diagrams, is given by
\begin{equation}
-2\pi g^{3}\delta((k-q)^2)k_{+}^{3}.
\label{50ff}
\end{equation}

As mentioned, we neglect contributions coming from parts $R_{1,2}$
in $V_{1,2}$, which are absent according to ~\cite{new}
and study the contributions coming from parts
$W_{1,2}$ only.

{\bf 1.} $(W_1)\cdot (A)^*$ (Fig. \ref{fig7}.1)

The product of reggeon propagators  and $R\to RP$ and $R\to RRP$
vertices is
\beq
ig^3{f^{ab_1d}f^{b_2cd}}f^{eb'_{1}c}
\frac{q_+}{(q-q_1)^2+i0}\bbb(p,q_2,q_1)\bbl(p,q'_1).
\eeq
The colour factor is
\begin{equation}
C_{1A}=-\frac{N_{c}^{2}-1}{32N_{c}}.
\label{66}
\end{equation}
The dependence on $q_{1-}$ comes from the denominators in
the vertex $W_{1}$:
\begin{equation}
D_{1}=(q-q_{1})^2+i0
\label{69}
\end{equation}
and the virtual quark propagator:
\begin{equation}
D_{2}=(k-q_{2}')^2-i0=(k+p-q+q_1-\lambda)^2-i0 .
\label{71}
\end{equation}
We get the longitudinal integral
\begin{equation}
I_{1A}=I(-q,k+p-q-\lambda)=0 .
\label{73}
\end{equation}

{\bf 2.} $(W_1)\cdot (C)^*$ (Fig \ref{fig7}.3)

The product of reggeon propagators  and $R\to RP$ and $R\to RRP$
vertices is
\beq
ig^3{f^{ab_1d}f^{b_2cd}}f^{eb'_{2}c}\frac{q_+}{(q-q_1)^2+i0}
\bbb(p,q_2,q_1)\bbl(p,q'_2).
\eeq
The colour factor is
\begin{equation}
C_{1C}=\frac{N_{c}^{2}-1}{16N_{c}}.
\label{105}
\end{equation}

The dependence on $q_{1-}$ comes from the denominators in the
vertex $W_{1}$ and the virtual quark propagator:
$D_1$ from (\ref{69}) and
\begin{equation}
D_{2}=(k-p-q_{2}')^2-i0=(k-q+q_{1}-\lambda)^2-i0 .
\end{equation}
We get the longitudinal integral
\begin{equation}
I_{1C}=I(-q,k-q-\lambda)=0 .
\label{110}
\end{equation}

{\bf 3.} $(W_2)\cdot (A)^*$ (Fig. \ref{fig7}.1)

The product of reggeon propagators and $R\to RP$ and $R\to RRP$
vertices is
\beq
ig^3{f^{ab_2d}f^{b_1cd}}f^{eb'_{1}c}\frac{q_+}{(q-q_2)^2+i0}
\bbb(p,q_1,q_2)\bbl(p,q'_1).
\eeq
The colour factor is
\begin{equation}
C_{2A}=-\frac{N_{c}^{2}-1}{16N_{c}}.
\label{c2a}
\end{equation}
The longitudinal integral is
\begin{equation}
I_{2A}=I(p,k+p-q-\lambda)=
\frac{2\pi i}{k_{+}q_{+}}
\frac{1}{\lambda_{-}+\frac{(p+q_1)^2_{\perp}}{q_{+}}
-\frac{q_{2\perp}^2}{k_{+}}+i0}.
\label{131}
\end{equation}
At both $k_+, p_+\to\infty$ this gives a non-zero contribution to the
inclusive cross-section.

{\bf 4.} $(W_2)\cdot (C)^*$ (Fig. \ref{fig7}.3)

The product of reggeon propagators  and $R\to RP$ and $R\to RRP$
vertices is
\beq
ig^3{f^{ab_2d}f^{b_1cd}}f^{eb'_{2}c}\frac{q_+}{(q-q_2)^2+i0}
\bbb(p,q_1,q_2)\bbl(p,q'_2).
\eeq
The colour factor is
\begin{equation}
C_{2C}=\frac{N_{c}^{2}-1}{32N_{c}}.
\label{c1c}
\end{equation}
The longitudinal integral is
\beq
I_{2C}=I(p,k-q-\lambda)=
\frac{2\pi i}{k_+q_+}
\frac{1}{\lambda_{-}+\frac{(p+q_1)_\perp^2}{q_+}
-\frac{(p+q_2)_\perp^2}{k_+}+i0} .
\eeq
At $k_+,p_+\to \infty$ we get a non-zero contribution to the inclusive
cross-section.

{\bf 5.}

The total contribution from interference terms between $(W_{1,2})$ and
$(A)+(B)+(C)+(D)$ is given by
\beq
(W_2)\cdot [(A)+(C)]^*+(\lambda\to -\lambda)+c.c.
\eeq
Supplying all the necessary factors we find the contribution to the
high-energy part
\beq
F_4=
-2\pi msg^{10}\frac{N_c^2-1}{N_c}\int\frac{d^2q_{1\perp}d^2q_{2\perp}}
{(2\pi)^4}\frac{1}{q_{1\perp}^4q_{2\perp}^4}
\bbb(p,q_2,q_1)\Big(\bbl(p,q_1)-2\bbl(p,q_2)\Big).
\eeq
It coincides with the contribution obtained in the BFKL-Bartels approach.

\section{Conclusions}

The main result of this paper is that after all longitudinal integrations
and with a certain restriction on the emitted gluon energy, the inclusive
cross-section for gluon production can be obtained directly from the
purely transverse BFKL-Bartels picture with the use of standard
Lipatov-Bartels vertices, as was done in~\cite{braun} on intuitive grounds.
This circumstance may be very important for calculation of inclusive
cross-sections in more complicated cases (such as nucleus-nucleus
collisions), where avoiding longitudinal integrations may lead
to substantial simplifications.
To get our result it has been essential that the R$\to$RRP vertex
depending on longitudinal variables and found by the direct application
of the effective action approach consists of two parts $W$ and $R$.
These two terms in their sum correctly describe the part of the
amplitude which does not contain $\delta$-functions in the
transferred "energies" $q_{1,2-}$. The latter are supplied by
the corresponding part of the contributions from the double
reggeon exchange. However, alternatively, one can take the
contribution from the double reggeon exchange as a whole,
substituting by it the part $R$ from the R$\to$RRP effective vertex.
Once this substitution has been done, the contribution
coming from the remaining part $W$ together with contributions from
the R$\to$RP vertices exactly coincides with what is found
in the transverse space BFKL-Bartels approach and thus fully confirms its
validity.

Note that in our paper ~\cite{bravyaz} this result required the principal
value prescription for the integration of certain singularities in
longitudinal integrals. In fact these singularities only appear in the
part $R$ of the R$\to$RRP vertex. So if one drops this part altogether the
principal value prescription, external with respect to the effective
action formalism, becomes redundant.

We finally note once again that the found contribution
is only a part of the total non-diffractive inclusive cross-section
The rest comes from intermediate states with only one of the targets
and requires knowledge of the R$\to$RRRP vertex. The study of the latter
is in progress.

\section{Acknowledgements}
The authors are indebted to J.Bartels and L.N.Lipatov for
helpful discussions. This work has been partially supported by
the RFFI grant 09-02-01327-a (Russia).


\begin{thebibliography}{99}
%
\bibitem{bfkl} L.N.Lipatov, Sov. J. Nucl. Phys. {\bf 23} (1976) 338;
\\ E.A.Kuraev, L.N.Lipatov and V.S.Fadin, Sov. Phys. JETP {\bf 45} (1977) 199;
\\ I.I.Balitsky and L.N.Lipatov, Sov. J. Nucl. Phys. {\bf 28} (1978) 822.
%
\bibitem{bartels} J.Bartels, Nucl. Phys. {\bf B175} (1980) 365.
%
\bibitem{lipatov} L.N.Lipatov, Phys. Rep. {\bf 286} (1997) 131.
%
\bibitem{bravyaz} M.A.Braun, M.I.Vyazovsky,
                 Eur. Phys. J. {\bf C 51} (2007) 103.
%
\bibitem{braun} M.A.Braun, Eur. Phys. J. {\bf C 48} (2006) 501.
%
\bibitem{new} M.A.Braun, L.N.Lipatov, M.Yu.Salykin and M.I.Vyazovski,
Eur. Phys. J. {\bf C 71} (2011) :1639.
%
\bibitem{gribov} V.N.Gribov, Sov. Phys. JETP {\bf 29} (1969) 483.
%
\bibitem{CK} A.Capella and A.Krzywicki, Phys. Rev. {\bf D 18} (1978) 3357.
%
\end{thebibliography}
\end{document}